\documentclass[aps,prl,preprint,tightenlines]{revtex4}
\usepackage{epsf}
\begin{document}
\preprint{UCHEP-08-02}
\title{Results from $\Upsilon$(5S) at Belle: \\
Strange Beauty and other Beasts\footnote{talk presented at the Lake Louise Winter Institute, 18-23 February, 2008}}

\author{K. Kinoshita}

\affiliation{University of Cincinnati \\
PO Box 210011 \\ 
Cincinnati, OH 45221, USA\\ 
E-mail: kay.kinoshita@uc.edu}

\begin{abstract}{
The Belle experiment collected in 2005-6 a total of 23.6~fb$^{-1}$ of data at the $\Upsilon(10860)$ resonance, also known as $\Upsilon$(5S).  
These constitute nearly all of the world's sample of $e^+e^- \to \Upsilon$(10860) events, which provide clean $B_s$ pairs.  
We present here several $B_s$ and $\Upsilon$(10860) properties recently extracted from these data.
}
\end{abstract}
\maketitle


Belle,\cite{belle} one of two $B$-factory experiments, is located at KEKB\cite{kekb} and studies primarily the $CP$ asymmetries of $B$ meson decay in $e^+e^-$ annihilations at the $\Upsilon$(4S) resonance ($Mc^2=10580\pm 1$~MeV, $\Gamma=20.5\pm 2.5$~MeV).\cite{pdg}
However, the richness of Belle physics reaches well beyond $CP$ (see, e.g., http://belle.kek.jp/);
an unequalled sample of $B\bar B$, $c\bar c$, $\tau\bar\tau$, and two-photon events enables access to many other topics. 
We now add $B_s$ and bottomonium studies, thanks to data collected at the $\Upsilon$(10860) resonance.

The $\Upsilon$(10860), ($Mc^2=10865\pm 8$~MeV/$c^2$,  $\Gamma=110\pm 13$~MeV),\cite{pdg} is interpreted as $\Upsilon$(5S), the fourth excitation of the vector bound state of $b\bar b$.
It is above $B_s\bar B_s$ threshold and, given the success of the $\Upsilon$(4S) program in characterizing properties of  $B_{d,u}$, it is natural to contemplate $B_s$ at $\Upsilon$(10860).
The $e^+e^-$ environment produces clean events, efficiently triggered,  with precisely known center-of-mass energy.
Furthermore, the $B$-factory offers an existing facility with high luminosity, a well-studied detector, and $\Upsilon$(4S) data for comparisons.
On the other hand, high integrated luminosities are needed,
$\sigma(e^+e^-\to \Upsilon$(10860))$\approx \sigma(e^+e^-\to \Upsilon$(4S)$)/3\approx 0.3$~nb, and events include $B_{d,u}$ as well as $B_s$.
While $B_s$ is produced copiously in hadronic collisions, the $\Upsilon$(10860) can be competitive, particularly in aspects of $B_s$ decay that are limited by systematic effects at a hadron machine.

Belle has run twice at the $\Upsilon$(10860).  
In June 2005 a three-day ``engineering'' run served to test KEKB, which had never operated at energies above the $\Upsilon$(4S), and study the basics of $\Upsilon$(10860), $B_s$, and $B_s^*$.
A scan of five energy points was used to locate the peak, at $\sqrt{s}=$10869~MeV, where 1.86~fb$^{-1}$ were collected.
By June 2006, results confirmed the projected potential of $\Upsilon$(10860), and 21.7~fb$^{-1}$ were collected in 20 days.

\begin{figure}[t]
\begin{center}
{\epsfysize=4.3cm\epsfbox{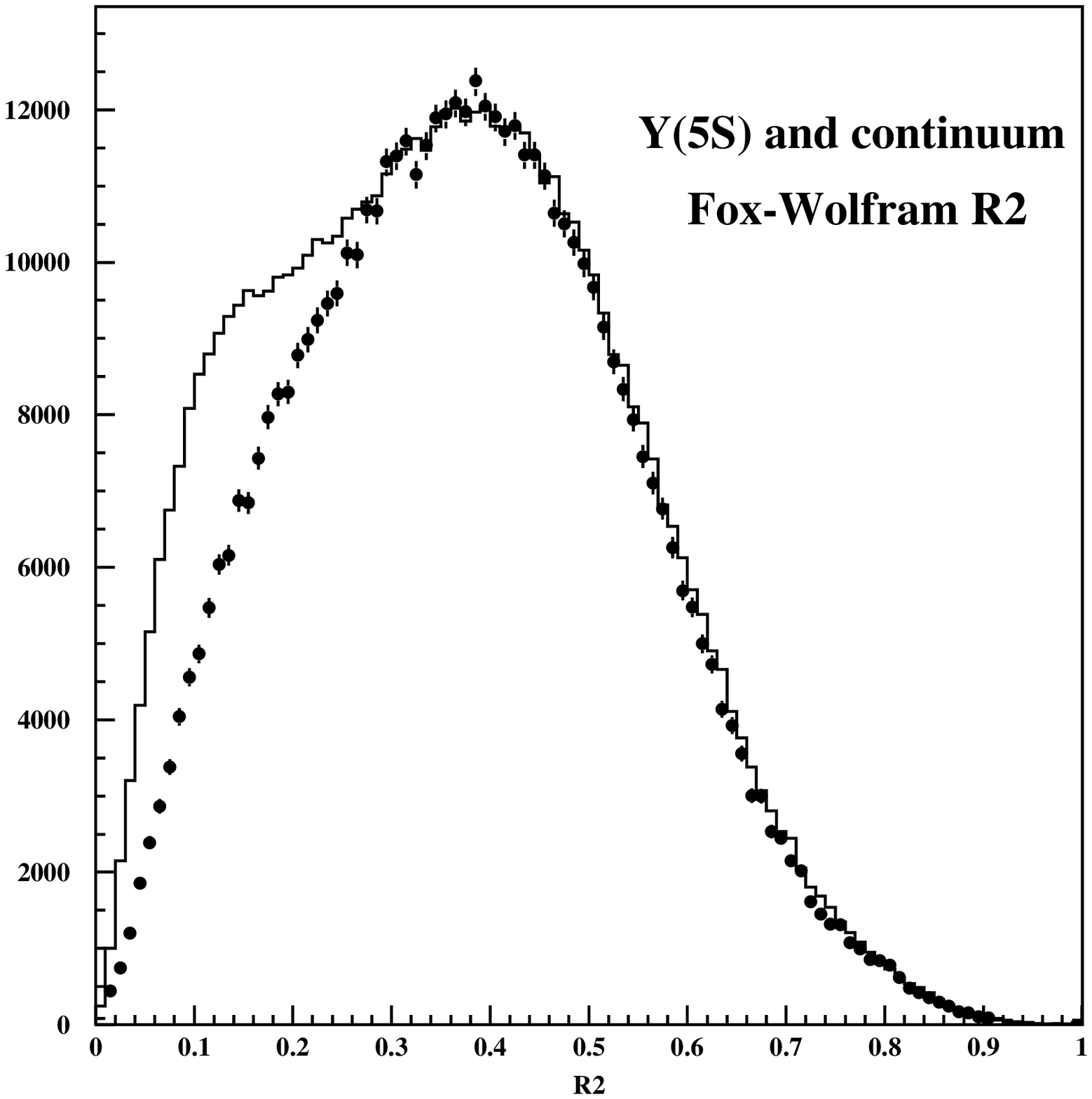}}~{\epsfysize=4.5cm\epsfbox{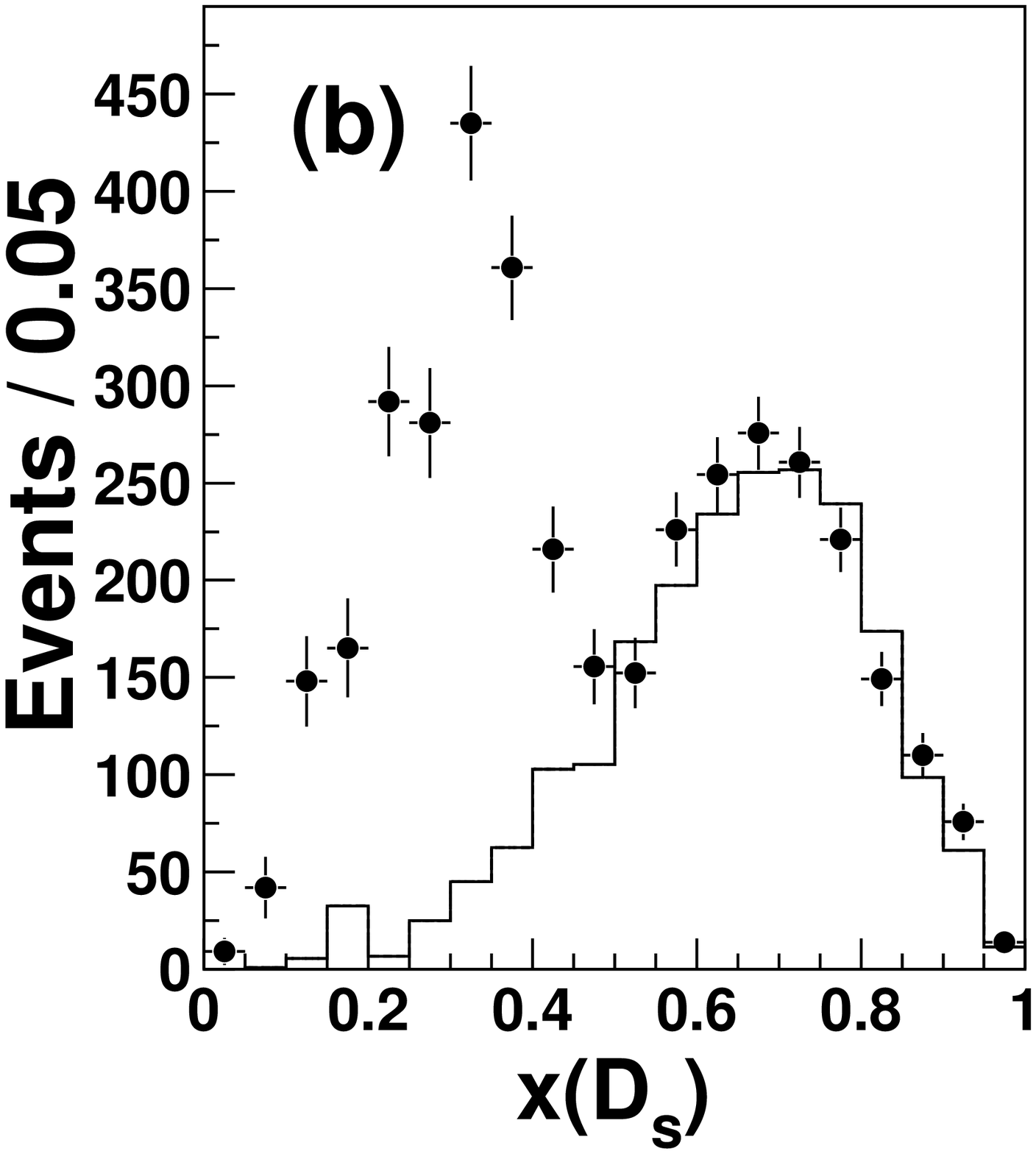}} 
\end{center} 
\caption{(Left) Distribution in $R_2$, (histogram) 1.86~fb$^{-1}$ at the $\Upsilon$(10860) and  (points) continuum below the $\Upsilon$(4S), scaled. (Right) Yield of $D_s$ mesons as a function of $x\equiv p_{D_s}/\sqrt{E_{beam}^2-M_{D_s}^2}$, (points) $\Upsilon$(10860) and (histogram) scaled  continuum.
 \label{5S_profile}}
\end{figure}

As prerequisite to studies of $B_s$, its abundance in $\Upsilon$(10860) events was determined from the 2005 data.\cite{5Sinclusive}
About 10\% of the hadronic events are resonance (assumed to be $b\bar b$), the rest being continuum $e^+e^-\to q\bar q$ ($q=u,d,s,c$).
Figure~\ref{5S_profile} (L) displays $R_2$, the ratio of the 2$^{nd}$ and 0$^{th}$ Fox-Wolfram moments,\cite{foxwolfram} a measure of ``jettiness'' that tends to be lower for the more isotropic resonance events.
We find $(3.01\pm0.02\pm 0.16)\times 10^5$~$b\bar b$~events/fb$^{-1}$.
The $b\bar b$ events may fragment to:
$B_s^{(*)} \bar B_s^{(*)}$, $B_q^{(*)} \bar B_q^{(*)}$, $B_q \bar B_q^{(*)}\pi$, $B_q \bar B_q\pi\pi$ ($q$ is a $u$- or $d$-quark).
To determine the fraction ($f_s$) that are $B_s^{(*)} \bar B_s^{(*)}$, we measure the inclusive rate ${\mathcal B}(\Upsilon(10860)\to D_s X)\equiv{\mathcal B}_{\Upsilon}$, an average over $B_s$, $B_d$, and $B_u$ (Figure~\ref{5S_profile}(R)).
Measured ${\mathcal B}(B_{u,d}\to D_s X)$ and our  understanding of decay mechanisms are used to estimate ${\mathcal B}(B_s\to D_s X)=(92\pm 11)\%$.\cite{cleo_5S}
Assuming that $B_d$ and $B_u$ are produced equally, ${\mathcal B}_{\Upsilon}$ depends only on $f_s$.
The same analysis, performed with inclusive $D^0$ yields, gives an independent value of $f_s$ with larger uncertainties;  ${\mathcal B}(B_s\to D^0X)\ll {\mathcal B}(B_q\to D^0X)$.
The combined result is $f_s=(18.0\pm 1.3\pm 3.2)\%$.\cite{5Sinclusive}

The results below, based on 23.6~fb$^{-1}$,  cover $B_s\to D_s\pi$, $D_s K$ (preliminary), $\Upsilon (10860)\to \Upsilon$(nS)$\pi^+\pi^-$,\cite{Ypipi} and $B_s\to\gamma\gamma$, $\phi\gamma$.\cite{gamgam}
$B_s$ decays are identified by ``full reconstruction'' of all final state particles, a method used with great success at the $\Upsilon$(4S).
Each candidate's energy and momentum in the $e^+e^-$ center-of-mass are evaluated as $\Delta E \equiv E_{cand}-E_{beam}$ and $M_{\rm bc}\equiv \sqrt{E_{beam}^2-p_{cand}^2}$.
In  $B_s\bar B_s$ (analogous to $\Upsilon$(4S)$\to B_q\bar B_q$), the $B_s$ carries the beam energy, so  $\left< \Delta E\right> =0$ and $\left< M_{\rm bc}\right> =M_{B_s}$.
For $B_s^*\bar B_s$ or $B_s^*\bar B_s^*$, the kinematics of $B_s^*\to B_s\gamma$ (with 50~MeV energy) leads to localized signals as well.
In the case of $B_s^*\bar B_s$ the result is effectively the loss of $50$~MeV from the $B_s$ pair, with the energy being shared approximately equally: $\left< \Delta E\right> \approx -25~$MeV and $\left< M_{\rm bc}\right>\approx M_{B_s}+25$~MeV.
For $B_s^*\bar B_s^*$ the energy reduction is $\sim$50~MeV.  

\begin{figure}[t]
\begin{center}
{\epsfxsize=5.5cm\epsfbox{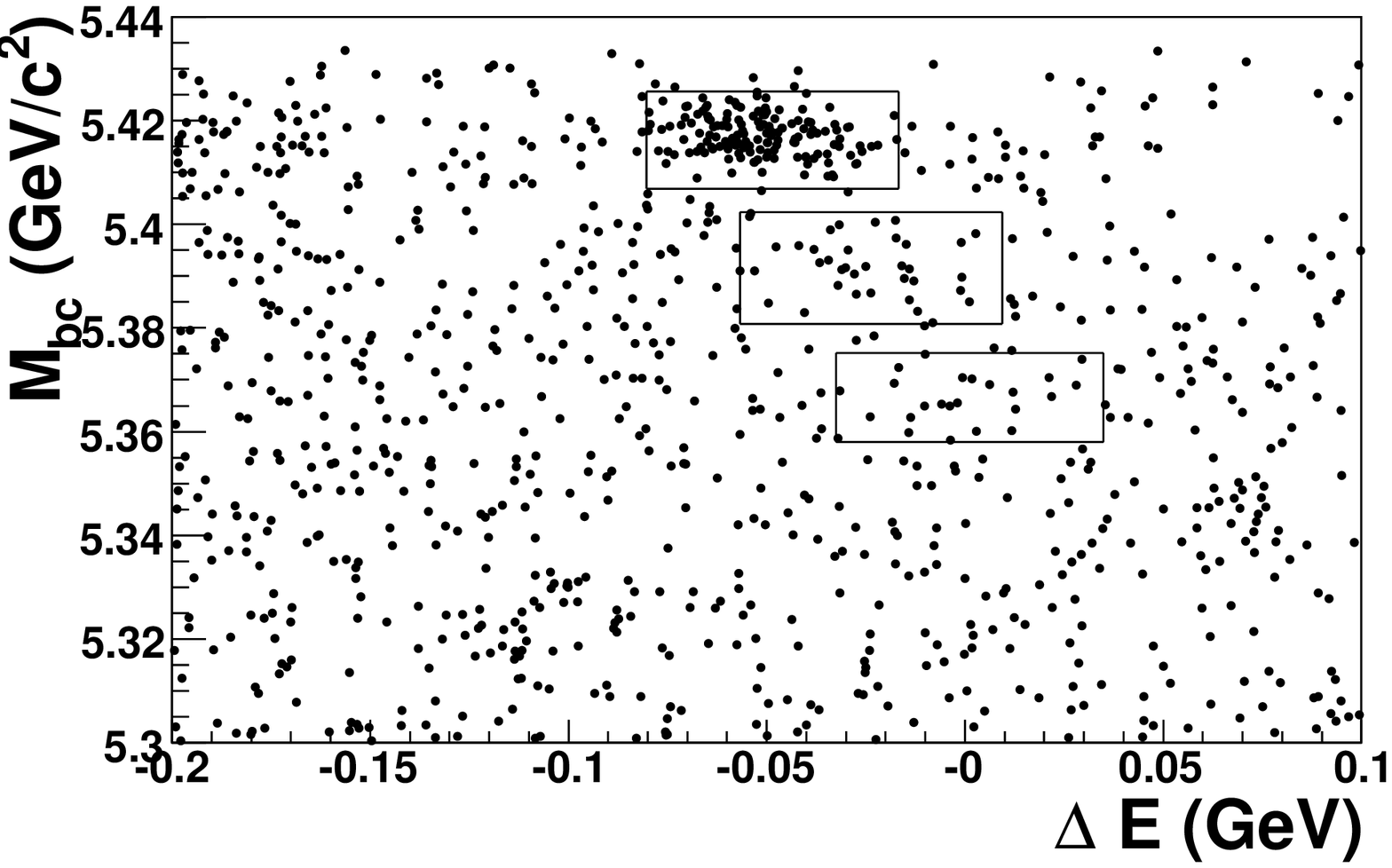}}{\epsfxsize=5.5cm\epsfbox{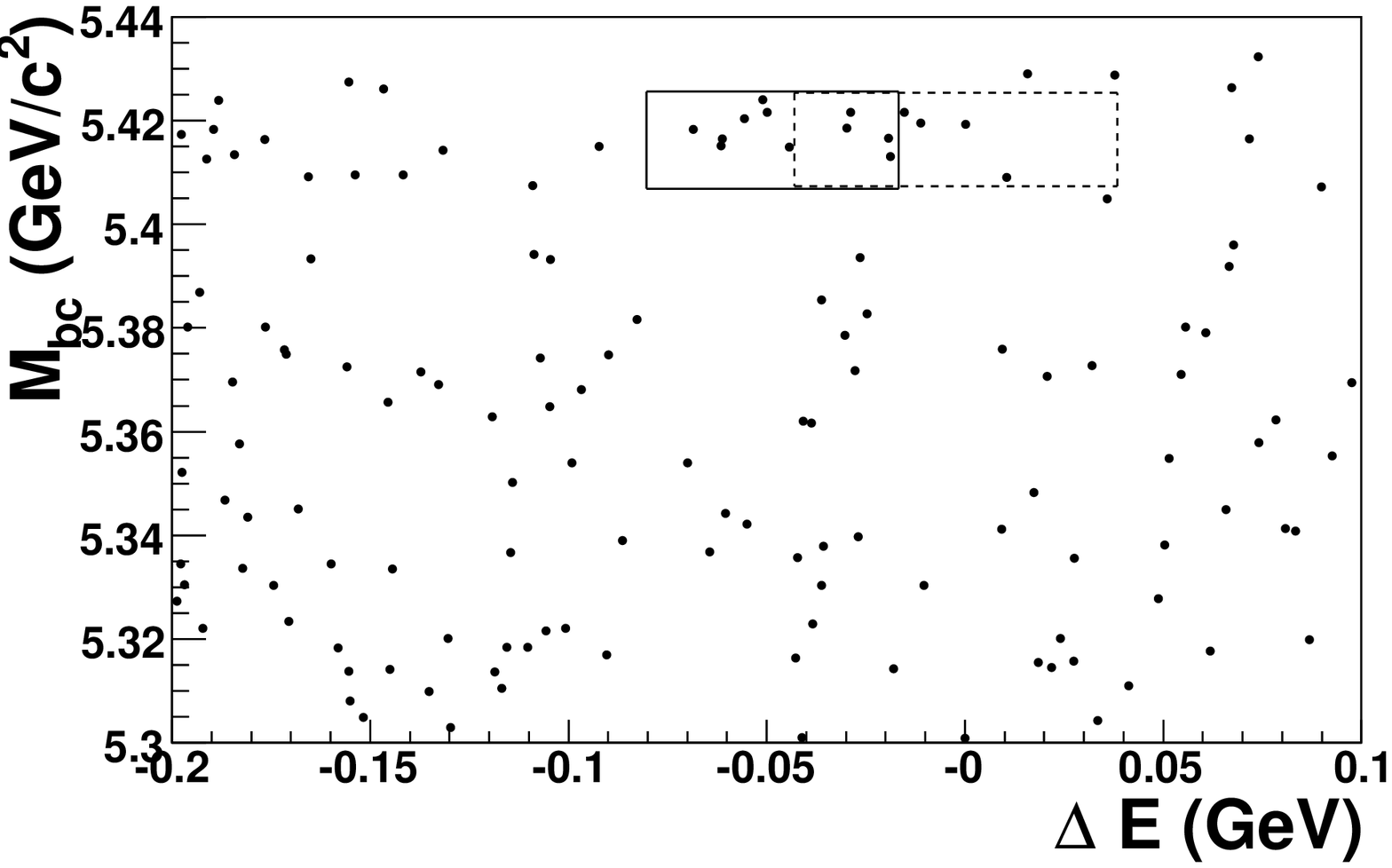}} 
\end{center} 
\caption{Distributions in $\Delta$E and $M_{\rm bc}$ of $B_s$ candidates  in 23.6~fb$^{-1}$ of Belle data at $\Upsilon$(10860). 
(Left) $B_s\to D_s\pi$, with  signal regions for  $B_s^*\bar B_s^*$(upper signal box), $B_s^*\bar B_s$(middle box), and $B_s\bar B_s$(lower box) events.
(Right)  $B_s\to D_s K$, with signal region for $B_s^*\bar B_s^*$  (solid box) and  region accumulating backgrounds from $D_s\pi$ (dashed box).
\label{Dspi}}
\end{figure}

The decay $B_s\to D_s^-\pi^+$  proceeds dominantly via a CKM-favored spectator process.
$D_s$ is reconstructed in 
$\phi (\to K^+K^-)\pi^-$, $K^{*0}(\to K^+K^-)K^-$, and $K_S(\to \pi^+\pi^-)K^-$. 
Shown in Figure~\ref{Dspi}(L) is the distribution in $\Delta E$ and $M_{\rm bc}$ for data, with signal boxes for $B_s^*\bar B_s^*$, $B_s^*\bar B_s$(inclusion of $B_s\bar B_s^*$ is implied), and $B_s\bar B_s$ events.
A two-dimensional fit yields $B_s\to D_s\pi$ signals of  $147^{+14}_{-13}$ ($20\sigma$), $12.7^{+6.0}_{-5.1}$ ($2.9\sigma$), and $3.8^{+4.5}_{-3.6}$ ($1.1\sigma$) in the $B_s^*\bar B_s^*$, $B_s^*\bar B_s$, and $B_s\bar B_s$ channels, respectively. 
Using $f_s=(19.5^{+3.0}_{-2.3})\%$ and $\sigma_{e^+e^-\to b\bar b}=0.302\pm 0.014$~nb (a weighted average from \cite{5Sinclusive,cleo_5Sb}), we obtain  
${\mathcal B}=(3.41^{+0.33}_{-0.31}({\rm stat.})^{+0.42}_{-0.41}({\rm sys.})^{+0.46}_{-0.45}({f_s})^{+0.33}_{-0.28}({{\mathcal B}(D_s^-\to\phi\pi)}))\times 10^{-3}$ and  $f_{B_s^*B_s^*}\equiv \frac{\sigma(e^+e^-\to B_s^*\bar B_s^*)}{ \sigma(e^+e^-\to B_s^{(*)}\bar B_s^{(*)})}
=(89.8^{+3.8}_{-4.0})\%$,  $f_{B_s^*B_s}\equiv \frac{\sigma(e^+e^-\to B_s^*\bar B_s + B_s\bar B_s^*)}{ \sigma(e^+e^-\to B_s^{(*)}\bar B_s^{(*)})}
=(7.8^{+3.3}_{-3.0})\%$.
The masses are found from the peak locations, $m_{B_s}=(5364.8\pm 1.3\pm 2.4){\rm MeV}/c^2$ and $m_{B_s^*}=(5417.6\pm 0.4\pm 0.5){\rm MeV}/c^2$.

The same process, CKM-suppressed, leads to $B_s\to D_s^-K^+$.
Given the lower expected rate, only $B_s^*\bar B_s^*$ is examined.  
Kinematic similarities result in a large incursion of $B_s\to D_s^-\pi^+$  into the signal region  (Figure~\ref{Dspi}(R)).
Good hadron identification is thus crucial.
A two-dimensional fit yields $6.8^{+3.4}_{-2.7}$ ($3.6\sigma$) events.
We obtain ${\mathcal B}=(2.2^{+1.1}_{-0.9}({\rm stat.})\pm 0.3({\rm sys.})\pm 0.3({f_s})\pm 0.2({{\mathcal B}(D_s^-\to\phi\pi)}))\times 10^{-4}$ and $\frac{{\mathcal B}(B_s\to D_sK)}{{\mathcal B}(B_s\to D_s\pi)}=(6.6^{+3.4}_{-2.8})\%$.

\begin{figure}[ht]
\begin{center}
{\makebox{\epsfxsize=5.5cm\epsfbox{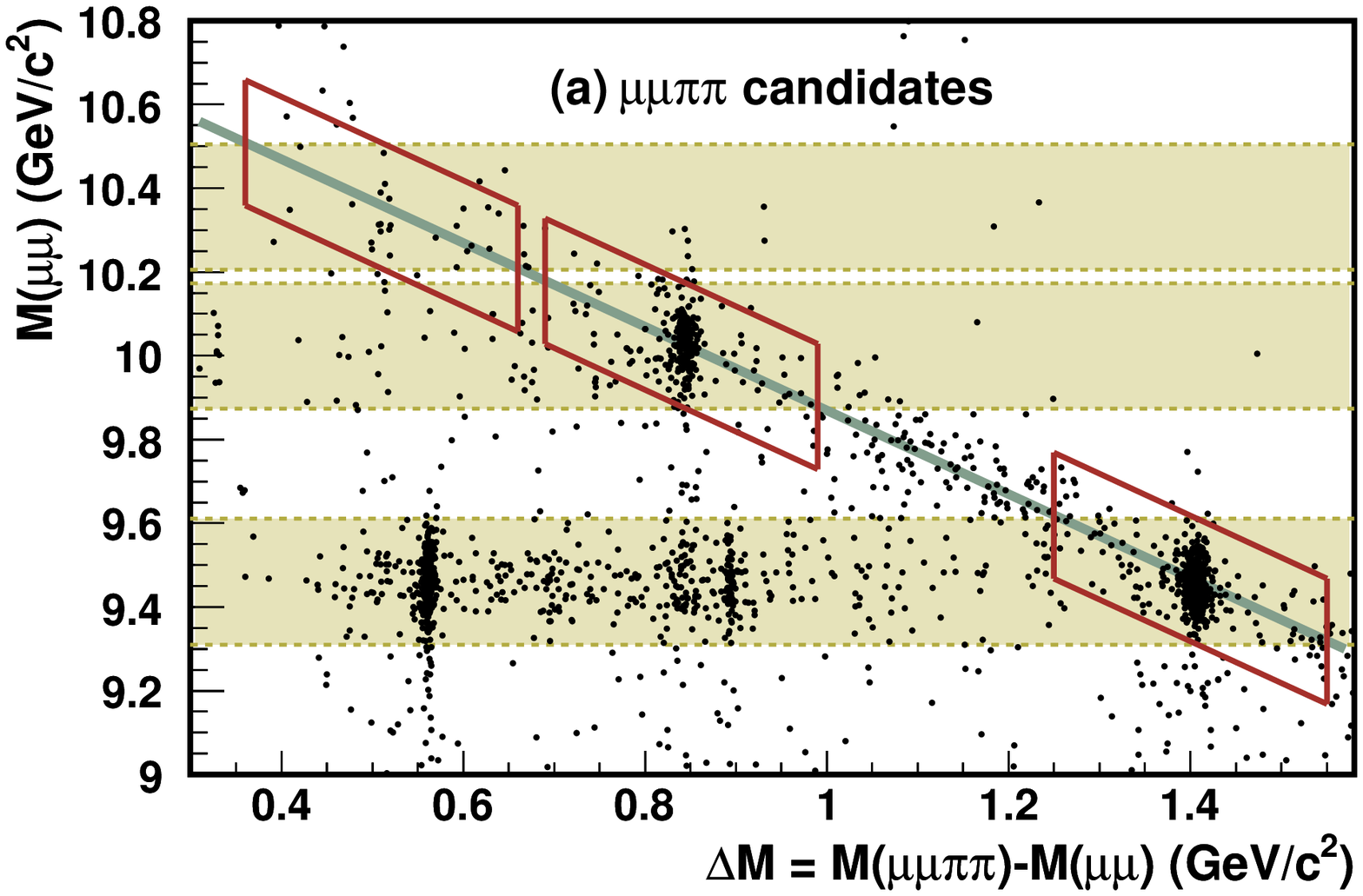}}
\makebox[5.5cm]{\vbox{\epsfxsize=5.5cm\epsfbox{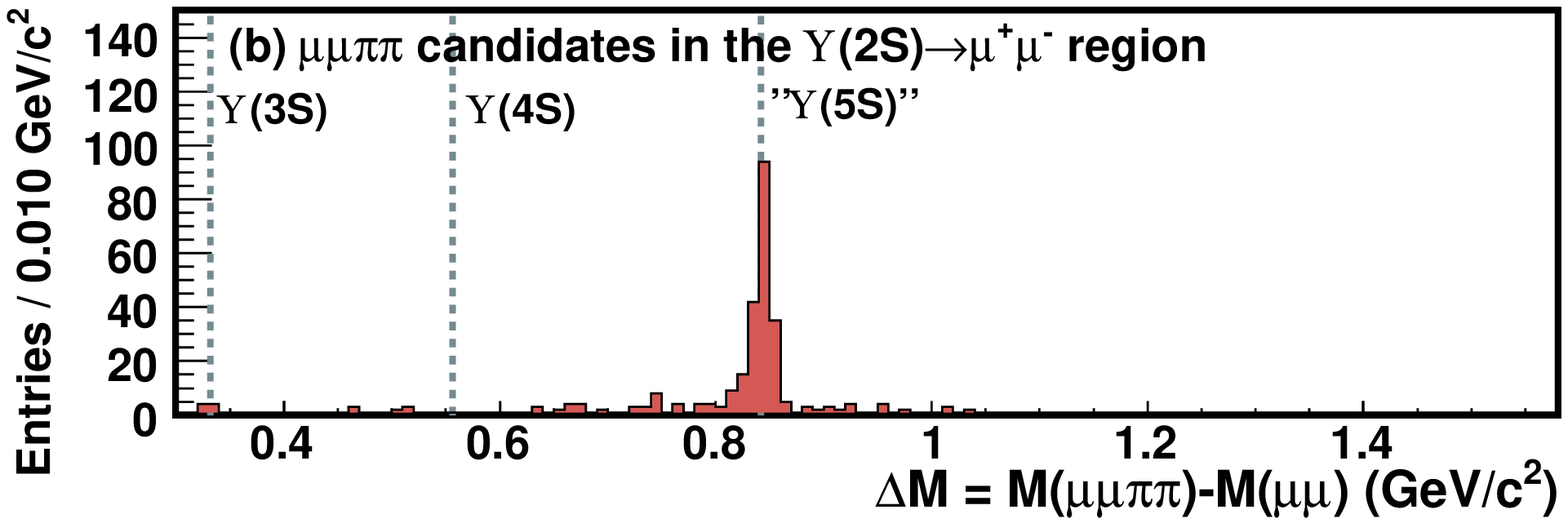}\\ \epsfxsize=5.5cm\epsfbox{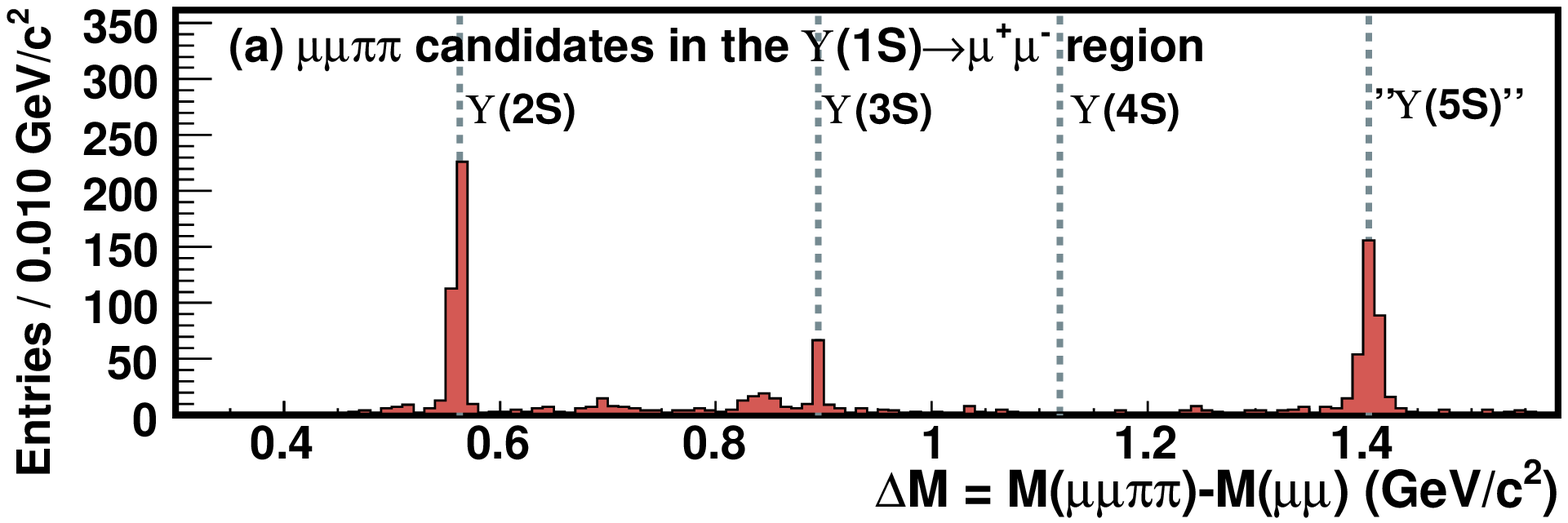}}}}
\end{center} 
\caption{(Left) Distribution in data of $M(\mu^+\mu^-)$ {\it vs.} $\Delta M\equiv M(\mu\mu\pi\pi)-M(\mu\mu)$ of $\mu^+\mu^-\pi^+\pi^-$ combinations. 
The line indicates the kinematic limit, $M(\mu\mu\pi\pi)=\sqrt{s}$. 
The horizontal bands indicate signal regions for (bottom to top) $\Upsilon$(1S), $\Upsilon$(2S), and $\Upsilon$(3S).
(Right) Projections onto $\Delta M$ of  (top)$\Upsilon$(2S) and (bottom) $\Upsilon$(1S) signal bands.
\label{fig:Ynspipi}}
\end{figure}

The observation of $Y(4260)$,\cite{Yc_first} and similar charmonium-like particles,\cite{Yc_next} raises the question of whether analogous bottomonium-like particles exist in the region near and above $\Upsilon$(10860).\cite{Yb_hou}
The $Y$'s are observed in modes $\psi(n$S$)h^+h^-$ where $h$ is $\pi$ or $K$; their nature is not yet fully understood.
Given widths of order 10~keV for $\Upsilon(n$S$)\to\Upsilon($1S)$)\pi^+\pi^-$  ($n=2,3,4$), one would na\"{i}vely expect ${\mathcal B}(\Upsilon($5S$)\to\Upsilon($1S$)\pi^+\pi^-)$ to be tiny, of order $10^{-5}$.  
Belle has searched for $\Upsilon$(10860)$\to\Upsilon(n$S)$(\to\mu^+\mu^-)h^+h^-$.  
Figure~\ref{fig:Ynspipi}(L) shows the distribution of  $\mu^+\mu^-$  mass and $\Delta M\equiv M(\mu\mu\pi\pi)-M(\mu\mu)$.
The horizontal bands indicate signal regions for $\Upsilon$(1S/2S/3S)$\to\mu\mu$.
Peaks correponding to $\Upsilon$(10860)$\to\Upsilon(n$S$)\pi^+\pi^-$ are seen in projections onto $\Delta M$ of $\Upsilon$ signal bands (Figure~\ref{fig:Ynspipi}(R)) which are fitted to determine signal yields.
From these we find cross sections, branching fractions, and partial widths for four modes, assuming $\Upsilon$(10860)=$\Upsilon$(5S): (in MeV)
$\Gamma(\Upsilon{\rm (5S)}\to\Upsilon{\rm (1S)}\pi^+\pi^-)=0.59\pm 0.04\pm 0.09$,  
$\Gamma(\Upsilon{\rm (5S)}\to\Upsilon{\rm (2S)}\pi^+\pi^-)=0.85\pm 0.07\pm 0.16$,  
$\Gamma(\Upsilon{\rm (5S)}\to\Upsilon{\rm (3S)}\pi^+\pi^-)=0.52^{+0.20}_{-0.17}\pm 0.10$,  
$\Gamma(\Upsilon{\rm (5S)}\to\Upsilon{\rm (1S)}K^+K^-)=0.067^{+0.017}_{-0.015}\pm 0.013$.
The $\Upsilon\pi\pi$ widths are of  order 1~MeV.

The widths for $\Upsilon$(2S/3S/4S/10860)$\to \Upsilon$(1S)$\pi^+\pi^-$\cite{pdg,Ypipi,Ypipi_other} are 0.0060, 0.0009, 0.0019,  and 0.59~MeV. 
The rate for $\Upsilon$(10860) is largest by two orders of magnitude.  
While the rate for  $\Upsilon$(5S) could be substantially enhanced,\cite{5S_enhance} our result raises the possibility that the $\Upsilon$(10860) is not a pure $\Upsilon$(5S) state.
To address this issue, Belle has scanned in energy  in the region above the $\Upsilon$(10860) resonance and anticipates first results by summer 2008.

Belle has searched for $B_s\to\gamma\gamma$ and $B_s\to\phi\gamma$.\cite{gamgam} 
Each includes a hard photon, readily isolated in the $B$-factory but difficult to trigger on at a hadron machine.  
For the Standard Model, ${\mathcal B }(B_s\to\gamma\gamma)$ is in the range $(0.4-1.0)\times 10^{-6}$;
extended models give rates up to $5\times 10^{-6}$.
The Belle $\Upsilon$(10860) data improve the limit 20-fold over previous searches, ${\mathcal B}<8.7\times 10^{-6}$ (90\%~$CL$).
The process for  $B_s\to\phi\gamma$ is identical to that for $B\to K^{(*)}\gamma$ (${\mathcal B}\sim 4\times 10^{-5}$) and is expected at this level.  
We report the first observation, ${\mathcal B}(B_s\to\phi\gamma)=(57^{+18}_{-15}(stat)^{+12}_{-11}(sys))\times 10^{-6}$, with a significance of 5.5$\sigma$.


In summary, Belle, which was designed for precision measurements of CKM parameters, has  opened a new avenue of inquiry by collecting 23.6~fb$^{-1}$ at the $\Upsilon$(10860) resonance.
The sample includes $\approx$1.3~million $B_s$ events and has yielded unprecedented sensitivity and precision for several modes.
In addition, an anomaly in the rate of $\Upsilon(10860)\to\Upsilon(n$S$)h^+h^-$ is stimulating new interest in bottomonium studies.

\section*{Acknowledgments}
The author wishes to thank the LLWI organizers and staff.
This work is supported by Department of Energy grant \# DE-FG02-84ER40153.

\end{document}